\documentclass{aa}

\usepackage{graphicx}
\usepackage{txfonts}
\begin{document}
\input{psfig.sty}

\title{Oxygen abundances in the Galactic bulge: evidence for fast chemical enrichment}

\author{
M. Zoccali\inst{1}
\and
A. Lecureur\inst{2}
\and
B. Barbuy\inst{3}
\and
V. Hill\inst{2}
\and
A. Renzini\inst{4}
\and
D. Minniti\inst{1}
\and
Y. Momany\inst{5}
\and
A. G\'omez\inst{2}
\and
S. Ortolani\inst{5}
\fnmsep
\thanks{Observations collected both at the European Southern Observatory,
Paranal, Chile (ESO programmes 71.B-0617 and 73.B-0074) and La Silla, Chile. } }
\offprints{M. Zoccali}
\institute{
P. Universidad Cat\'olica de Chile, Departamento de Astronom\'\i a y 
Astrof\'\i sica, Casilla 306, Santiago 22, Chile;\\
e-mail: mzoccali@astro.puc.cl, dante@astro.puc.cl
\and  
Observatoire de Paris-Meudon,  92195 Meudon Cedex, France;\\
e-mail: Vanessa.Hill@obspm.fr, Aurelie.Lecureur@obspm.fr, Ana.Gomez@obspm.fr
\and
Universidade de S\~ao Paulo, IAG, Rua do Mat\~ao 1226,
Cidade Universit\'aria, S\~ao Paulo 05508-900, Brazil;\\
e-mail: barbuy@astro.iag.usp.br
\and
INAF - Osservatorio Astronomico di Padova, Vicolo dell'Osservatorio 2,
I-35122 Padova, Italy; \\
e-mail: renzini@pd.astro.it
\and
Universit\`a di Padova, Dipartimento di Astronomia, Vicolo
dell'Osservatorio 5, I-35122 Padova, Italy;\\
e-mail: momany@pd.astro.it, ortolani@pd.astro.it
}

\date{Received: May 25, 2006; accepted: July 16, 2006}
                                                              
\abstract
{}
{We spectroscopically  characterize  the Galactic  Bulge to  infer its
star formation timescale,  compared to the other Galactic  components,
through the chemical signature on its individual stars.}
{We derived iron and oxygen abundances for  50 K giants in four fields
towards the Galactic bulge. High resolution (R=45,000) spectra for the
target stars were collected with FLAMES-UVES at the VLT.}
{Oxygen, as measured  from the forbidden line at   6300 \AA , shows  a
well-defined trend with [Fe/H], with [O/Fe] higher in bulge stars than
in thick disk ones,  which were known to  be more oxygen enhanced than
thin disk stars.}
{These results support a scenario in which the bulge formed before and
more rapidly than the disk, and therefore the MW bulge can be regarded
as a  prototypical old spheroid, with  a  formation history similar to
that of early-type (elliptical) galaxies.}

\keywords{Galaxy: bulge - Stars: Abundances, Atmospheres}
\titlerunning{Oxygen in the Galactic bulge}
\authorrunning{M. Zoccali et al.}

\maketitle


\section{Introduction}

The central  regions of spiral galaxies,  known as bulges, are made of
stars  in randomly  oriented  orbits for  which two distinct formation
processes have  been proposed.  In  the so-called ``classical bulges",
most stars originate  in  a short  phase  of star formation   when the
universe   was only   a   few Gyr  old.    Instead,  in  the so-called
``pseudobulges", stars form in the disk over a more extended period of
time, and  the bulge  results from the  secular evolution  of the disk
driven by the development of a bar (Kormendy \& Kennicutt 2004). It is
currently  believed that the  pseudobulge  mode dominates in late-type
spirals (Sc  and  later) and  the old-starburst  mode (similar to  the
formation of elliptical galaxies) dominates  in early-type spirals (Sb
and earlier). The Sbc Milky Way galaxy sits  on the borderline, and to
establish   the origin of its    bulge (either old   starburst or disk
secular  evolution),  extensive  spectroscopic observations have  been
undertaken to obtain precise  element abundance ratios, in  particular
[Fe/H] and [O/Fe].

The detailed chemical composition   of stars carries the signature  of
the enrichment  processes undergone by  the interstellar  medium up to
the moment of their formation. Thus, elemental ratios are sensitive to
the  previous  history of  star formation   and  can be used  to infer
whether there is a genetic link  between different stellar groups.  In
particular, the relative abundances of iron and $\alpha$-elements play
a  key r\^ole because $\alpha$-elements  are predominantly produced by
Type II supernovae (SNII) while  supernovae of Type Ia (SNIa) dominate
instead  the     production of  iron. The  SNIIs    come from massive,
short-lived stars, while SNIas result   from binary evolution and  are
characterized by a very broad distribution  of delay times (from a few
$10^7$ to over $10^{10}$ yr) between  the star formation event and the
SN explosion (Greggio 2005). As a consequence, the [$\alpha$/Fe] ratio
depends on the relative contribution of SNIIs and SNIas, and therefore
it  depends on the  timescale of  star-formation and metal  production
(Matteucci \& Greggio 1986).

Among  the $\alpha$-elements, oxygen   is especially important because
produced only by SNII's, and  its forbidden line  [OI] at 6300 \AA\ is
easily measured on  the spectra of  red giant stars. This line  allows
one to derive reliable  abundances  because its atomic parameters  are
well known; and since it is due to  a transition from the ground state
to  a  collisionally controlled higher  level,   it is not  subject to
non-LTE effects (Asplund et al.  2004).   Based on this line, accurate
measurements of the oxygen  abundance in the  Galactic thick  and thin
disk are  also available (Bensby et al.   2004), while several efforts
have been made to measure detailed abundances  in the bulge (McWilliam
\& Rich 1994,  2003; Rich \&  Origlia 2005).  However,  the bulge data
were   rather scanty,  and   the   resulting   evidence  was  somewhat
contradictory. To overcome these limitations, high quality spectra for
a large sample of bulge stars are presented.


\section{Observations and data analysis}

\begin{table}[ht!]
\caption{Characteristics of the four bulge fields}
\label{fields}
\begin{tabular}{clcccc}
\hline\hline
\noalign{\smallskip}
Nr.  &  Identification   &  $l$ & $b$ & R$_{\rm GC}$ & E$(B-V)$ \\
     &                   &      &     &   (pc)       &          \\
\hline
   1  & $b=-6$ Field   &  0.21 & $-6.02$ &  850 & 0.48   \\
   2  & Baade's Window &  1.14 & $-4.18$ &  604 & 0.55   \\
   3  & Blanco $b=-12$ &  0.00 & $-12.0$ & 1663 & 0.20   \\
   4  & NGC~6553 Field &  5.25 & $-3.02$ &  844 & 0.70   \\
\hline
\end{tabular}
\end{table}

Spectra for a sample of $\sim1000$ K giants in  four bulge fields have
been collected at the VLT-UT2 with the FLAMES fibre spectrograph.  All
the stars were observed  with the GIRAFFE  arm of the instrument, with
resolution R$\sim20,000$, while    58 of them  have {\it   also}  been
observed with the UVES arm, at higher resolution R$\sim45,000$, in the
range  5800-6800  \AA.  A complete description  of  the  whole GIRAFFE
sample will be given elsewhere  (Zoccali et al. 2006, in preparation),
here we focus  on the high-resolution UVES spectra,  for which the S/N
$\sim$50 per resolution   element has allowed  us  to obtain  accurate
oxygen abundances.  The bluer half of the UVES  spectrum has lower S/N
and was only  used to derive  carbon abundances  from a  C$_2$ band at
5635 \AA.  Observed targets include 11 stars in a low reddening window
at $(l,b)=(0,-6)$, 21  stars in Baade's Window,  5 stars in the Blanco
field at  $(l,b)=(0,-12)$, and 13 stars in  a field near  the globular
cluster   NGC~6553 (Table~\ref{fields}).   In    the  colour-magnitude
diagram, these stars  are  located on the red  giant  branch,  about 1
magnitude  above  the  red clump, with  the  exception  of 13 stars in
Baade's Window that instead are on the red clump itself.

Individual    spectra  were  reduced    with the standard  FLAMES-UVES
pipeline,    including   bias,    flatfield,   interorder   background
subtraction,    extraction,   wavelength  calibration,    and    order
merging. All the spectra  for each star   (a number between 7  and 17,
depending on the field) were then  registered in wavelength to correct
for heliocentric radial velocity and   averaged (rejecting the  lowest
and highest values) to a single spectrum per star.  In each plate, one
UVES fibre was allocated  to an empty  sky region.  This spectrum  was
registered  in flux to match the  equivalent width of the sky emission
lines of  each  object  spectrum and  then   subtracted from it.   The
spectrum  of a fast-rotating  B   star, thus containing only  telluric
absorption   lines, was   coadded to  itself    using different radial
velocity shifts, reproducing the shifts  applied to the target spectra
to remove the heliocentric, variable component of the radial velocity.
Stars with telluric absorption blended with  the O line were rejected.
Stars  with such absorption  in the wings of  the O line were included
with a different symbol.  The strong sky oxygen emission line prevents
a measurement of  the oxygen stellar  line when the radial velocity of
the  star is close to zero,  and these stars  were eliminated from the
sample.  In the end, 8 stars were discarded, leaving us with 50 stars.

All the stars have $V,I$  magnitudes, obtained either from  photometry
performed on  wide field WFI images from  the ESO 2.2m telescope at La
Silla (fields 1,3,4) or from the  OGLE catalogue (Udalski et al. 2002)
(field 2).  In addition, $J,H,K_s$ magnitudes are available for all of
them from  the 2MASS point source catalogue  (Carpenter  et al. 2001).
The assumed reddening for  each field is listed in Table~\ref{fields},
while standard extinction laws (Rieke  \& Lebofski 1985) were used  to
derive  extinctions  in   the  other   bands.  The   five  de-reddened
magnitudes  were  combined  to   obtain photometric  temperatures from
$V-I$, $V-J$,  $V-H$,  and  $V-K$  colours, according  to  the  latest
empirical calibration (Ram\'{\i}rez \&  Mel\'endez 2005). The  mean of
the four values was used as  a first guess for spectroscopic analysis.
Photometric gravity was instead calculated from the classical relation
\[
\log g_*= \log g_\odot+4\log \frac{T_*}{T_\odot}+0.4(M_{\rm bol}-M_{\rm bol \odot})+\log \frac{M_*}{M_{\odot}}
\]
by adopting a mean distance of $8$ kpc for the bulge, T$_{\odot}$=5770
K, $\log g_{\odot}$=4.44,  M$_{\rm  bol \odot}$=4.75,  and  M$_*$=0.85
M$_{\odot}$.

The equivalent widths  for selected lines of  Fe, Na, Mg, Al,  Si, Ca,
Sc,  Ti, and Ni  were measured  using  the new automatic code  DAOSPEC
(Stetson and Pancino in prep.).  The selection  of clean Fe lines, and
their atomic parameters was compiled using a spectrum  of $\mu$ Leo as
reference, observed   at the  Canada-France-Hawaii Telescope  with the
ESPaDOnS spectrograph, at resolution R=80,000  and S/N$\sim 500$.  The
following parameters were determined for $\mu$ Leo: T$_{\rm eff}$=4550
K, logg=2.3, microturbulence velocity $V_{\rm t}$=1.3 km/s . Following
Fulbright et al.  (2006) we carried on  our analysis {\it relative} to
a nearby reference star; i.e., we derived relative log gfs using $\mu$
Leo as a reference, requiring [Fe/H]=+0.30  for each single line. More
details on the  analysis of $\mu$ Leo,  the selection of clean  lines,
and the choice of its metallicity are given  in Lecureur \& al (2006).
With the same atomic parameters, we obtain [Fe/H]=$-0.52$ for Arcturus
and [Fe/H]=$0.04$ ($\epsilon$(Fe)$_\odot$=7.55) for the Sun.  This was
then included as  zero point for the iron  determination of our target
stars.  Atomic parameters for the [OI] and  NiI lines around 6300 {\rm
\AA}  were  instead assumed  to  be identical  to   those employed  in
previous measurements   for disk stars,  so  as to allow  a meaningful
comparison  between bulge  and   disk abundance patterns  (Table~2  in
Bensby et al.  2004). For reference, the method outlined below returns
[O/Fe]=$-0.01$  and  [O/Fe]=$+0.05$   for  the   Sun and  $\mu$   Leo,
respectively.  The damping constants were computed where possible, and
in particular    for most of  the  FeI   lines, using the  collisional
broadening theory (Barklem,  Anstee \& O'Mara 1998,  Barklem, Piskunov
\& O'Mara 2000).

The LTE abundance analysis was  performed using well tested procedures
(Spite 1967)  and  the new MARCS  models  (Gustafsson  et  al.  2002).
Spectrum synthesis was performed with {\it turbospec} (Alvarez \& Plez
1998) and  counterchecked with Barbuy  et   al. (2003), including  the
effects of molecular lines on the derived  atomic abundances, which is
of special importance in the  case of oxygen,  due to the formation of
CO molecules, locking part of the  oxygen, and TiO molecules, changing
the continuum  shape. Excitation equilibrium   was imposed on the  FeI
lines    in order to  refine  the   photometric  T$_{\rm eff}$,  while
photometric gravity was imposed even if ionization equilibrium was not
fulfilled.  The reason for this choice is that FeII lines are all very
weak and  contaminated by CN. On the  other hand, while the strong and
differential reddening has a strong effect  on the photometric T$_{\rm
eff}$   (why  we prefer the    excitation one), photometric gravity is
robust.     In  fact, an error  of   $\Delta  E(B-V)=0.05$  implies an
uncertainty of   $\Delta$  T$_{\rm  eff}=100K$  in  the    photometric
temperature but only $\Delta$logg=0.015 in the photometric gravity.

A synthetic spectrum was calculated for each star, with the parameters
and Fe  given above, adopting  solar abundance (see  below) as a first
guess and  then iterating until  the oxygen line was  reproduced well.
Other elements   whose abundance  might affect  the  oxygen  were also
measured   in  the  iterations, namely:   carbon   from the C$_2$(0,1)
bandhead  of  the  Swan  A$^3\Pi_g$-X$^3\Pi_u$  system  at 5635   \AA,
nitrogen from the CN band at 6498.5 \AA , Ti, and Ni. The [C/Fe] ratio
was  found to range  between  $-$0.30 and  +0.10  for our  stars.  The
accuracy we have  in these measurements  is  not very high ($\pm  0.2$
dex), but we  verified that a  variation of  {\bf $+ 0.2$}  dex in the
adopted C  implies a  negligible variation in  [O/Fe], except  for the
coolest stars at super-solar [Fe/H], for which it  can introduce up to
a 0.1 increase in [O/Fe].  Nitrogen, on the other hand, ranges between
[N/Fe]=0.00  and [N/Fe]=+0.50. Particular  care was taken in measuring
Nickel, given  that the [OI]  6300.304 {\rm \AA}  line is blended with
the  two  isotopic components  of the NiI  line  at 6300.335 {\rm \AA}
(Johansson et al.  2003).  Thus isolated Ni lines were used to measure
[Ni/Fe], and this value was assumed in the  deblending of the [OI]6300
{\rm \AA} line. Standard  solar abundances  (Grevesse \& Sauval  1998)
were adopted as  reference zero   point  for the abundances  of  bulge
stars,   except    for     the  value    of      solar oxygen,   where
$\epsilon$(O)$_\odot$ = 8.77 was assumed  (Allende Prieto, Lambert  \&
Asplund 2001).

From the dispersion  of Fe lines, we estimate  a  statistical error of
0.05  dex on Fe. On the  other hand, a  systematic error of $-$100K in
T$_{\rm eff}$ implies $\Delta [Fe/H]=-0.04$ and a negligible effect on
[O/H], while $\Delta [Fe/H]=+0.04$ and a $\Delta [Fe/H]=+0.1$ would be
given by a +0.3 dex increase in log g.

A table with the complete list of parameters and abundances, including
sodium, magnesium, and  aluminum for each star will  be published in a
companion paper (Lecureur et al. 2006).


\section{[O/Fe] ratios}

\begin{figure}[ht]
\psfig{file=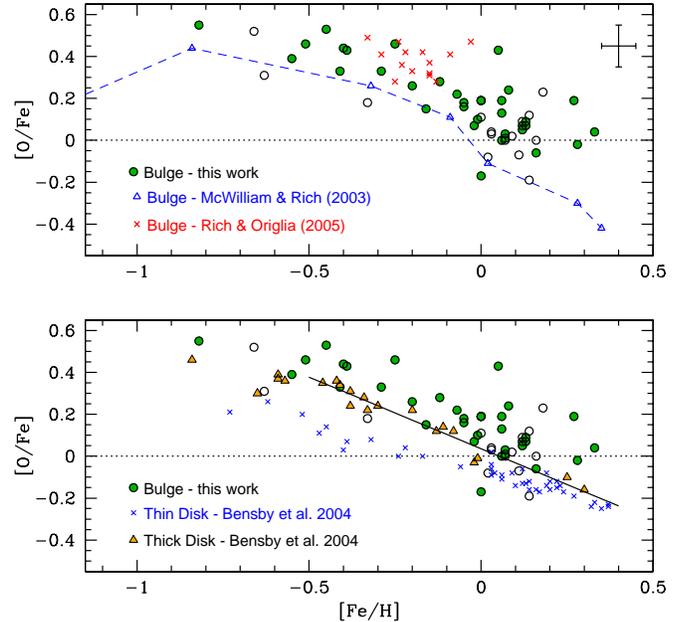,angle=0,width=9.0 cm}
\caption{
{\bf Upper panel:} The [O/Fe] vs. [Fe/H for our bulge stars (circles)
along  with previous determinations  in other bulge stars from optical
(open  triangles) and near-IR spectra (crosses).   Open symbols in our
measurements  refer to  spectra  with lower S/N  or  with  the O line
partially blended  with  telluric  absorption.   {\bf  Lower   panel:}
oxygen/iron trend in our bulge stars vs.  that for thick and thin disk
stars.  The  solid line  shows a linear   fit to  the thick  disk data
points with [Fe/H]$>-0.5$  and is meant  to emphasize  that all bulge
stars  with $-0.4<$[Fe/H]$<+0.1$  are more oxygen-enhanced than thick
disk stars.  This  plot enlightens the  systematic, genetic difference
between bulge and disk stars,   thus excluding that bulge stars   were
once disk stars that then migrated inward to build up the bulge.  }
\label{oxy}
\end{figure}

The resulting [O/Fe] vs.  [Fe/H] plot is shown in Fig.~\ref{oxy}.  The
upper panel shows the comparison  between our measurements and  recent
oxygen abundance determinations for bulge stars.  Earlier measurements
(McWilliam \&  Rich 1994) are  omitted here as  more  uncertain due to
their much lower spectral  resolution.   A few corrections have   then
been applied to ensure   consistency between our bulge  abundances and
those from previous determinations for bulge  and disk stars.  For the
bulge optical  data (McWilliam \&  Rich 2003), a  shift  of $-0.06$ in
[O/Fe] takes the difference in the adopted solar oxygen abundance into
account,  their $\epsilon$(O)$_\odot$=8.71 versus   our 8.77,  while a
further +0.1 in [O/Fe] corrects for the different log$gf$=$-9.717$ for
the [OI]  line, 0.1 dex  lower than the  one adopted here.  Similarly,
the bulge abundances from near-IR  measurements (Rich \& Origlia 2005)
were shifted upwards  by +0.06, due  to their  assumption of a  higher
solar oxygen,  $\epsilon$(O)$_\odot$=8.83.  This figure shows that our
measurements  do  not conflict with  previous  results, but the larger
size and metallicity coverage of the present sample  allows us to draw
much more robust conclusions.

The lower  panel of Fig.~\ref{oxy} shows the  {\it main result} of the
present   investigation: the [O/Fe]  vs.   [Fe/H]  ratios in the bulge
compared  with those  for  the thick   and the thin   disks (Bensby et
al. 2004).   These measurements are  as consistent as possible, in the
sense that they  come from the   same line, and we have  intentionally
adopted the   same atomic parameters  both for  oxygen and for nickel.
The solar  oxygen abundance to which the  disk stars were referred was
lower by 0.06 dex, so an identical downward  shift was then applied to
their measurements.  This plot shows that  the thin disk,  thick disk,
and  bulge evolved through  different chemical trajectories.  In other
words,  bulge stars did  not  originate in the   disk and then migrate
inward to  build up the bulge,  but rather formed independently of the
disk  (Minniti  1995,  Ortolani et al.    1995).   Moreover,  {\it the
chemical enrichment  of the bulge, hence  its formation timescale, has
been faster than that of the thick disk, which in turn was faster than
that of the thin disk (Matteucci, Romano
\& Molaro 1999).}

A  few stars in  our sample show lower  [O/Fe] than the  mean locus of
thick disk stars.  Their number is consistent with the expected number
of  foreground disk contaminants.  According  to  the Besan\c on model
(Robin et al. 2003)  and  to the  disk  control  field in  Zoccali  et
al. (2003) the   contamination fraction is  15$\%$  in all our  fields
except the Blanco one,  where it is  45$\%$.  All except two of  these
stars would indeed     require  stronger gravity  according  to    the
ionization equilibrium.  Two extreme  cases (logg$\sim$3-4)  have been
excluded  from the sample  so are  not shown  here.  We also  excluded
stars with radial velocity close to zero (with a  sky emission line on
top of   the oxygen line), further  biasing  our  sample  against disk
contaminants.   We therefore conclude that  the stars  above the thick
disk line in Fig.~1 are likely to be bona fide bulge stars.
 
It is  evident  from Fig.~1 that the  dispersion  in [O/Fe]  for bulge
stars is higher  than for both thin   and thick disks. While  at least
part  of this effect  is expected, given  that  disk stars are nearby,
bright and isolated stars for which  higher S/N spectra were obtained,
the possibility  that  part of  the    dispersion is real   cannot  be
excluded, given that in  the bulge we are sampling  stars in a  region
much larger than the solar neighbourhood.  It is worth mentioning that
no [O/Fe] gradient was present among the four different fields.

\section{Conclusions}

Age-dating of the bulge  stars from their colour-magnitude diagram has
already demonstrated that the  bulk of them  are older than $\sim  10$
Gyr, with no  detectable trace of intermediate  age stars (Ortolani et
al.  1995; Kuijken \& Rich 2002; Zoccali et al.  2003). Interpretation
of the evidence  now  provided  on the    [O/Fe] ratios is   that  the
formation of the Galactic bulge was  faster than that  of the disk and
preceeded it.  The most metal-rich bulge  stars were the latest formed
during  the bulge   evolution,   and  their  ages correspond   to  the
exhaustion  of interstellar matter in the  bulge.  The present results
reveal that  the oxygen-to-iron starts to drop  at the  metallicity of
[Fe/H]$\approx$-0.2 due to the progressive enrichment by SNIa.  A fast
star formation rate is needed to explain this result.

\begin{figure}[ht]
\psfig{file=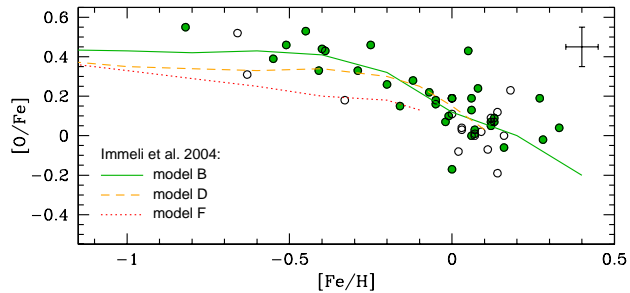,angle=0,width=8.5 cm}
\caption{Comparison between the observed [O/Fe] ratio of bulge stars
and the prediction of the simulations by Immeli et al. (2004). See text
for details.}
\label{immeli}
\end{figure}

It should be noted that we have compared the elemental ratios of bulge
stars to those of disk stars in the {\it solar neighborhood}. Strictly
speaking, our result implies that the bulge stars did not originate in
disk material  (either already stellar or  gaseous) coming from as far
as   the solar neighbourhood. Due to   the possible presence of radial
gradients in the disk, a comparison with stars in the inner disk would
have been more conclusive.  Unfortunately no  such sample is available
at  the  moment.  An  alternative approach  to  interpreting  of these
results is  the    comparison with   theoretical  simulations.      In
Fig.~\ref{immeli} we compare our results with the predictions of bulge
[O/Fe] ratios by Immeli et  al.  (2004).  They simulate the  formation
of galaxies from clouds with different  dissipation efficiencies.  The
case with high dissipation (model B) rapidly forms massive clumps that
spiral to  the  centre and merge to   a central  bulge component  in a
strong starburst.  In contrast, when the dissipation efficiency is low
(model F) the galaxy  forms a thicker disk,  and  only at  later times
does instability  set in, driving the  formation  of a bar and  then a
pseudobulge.  Model D is the intermediate case.  Clearly, our data can
exclude model F,  while it agrees with  model B very well.  This makes
the MW bulge similar to early-type galaxies, in being $\alpha$-element
enhanced, dominated by old stellar populations, and having formed on a
timescale shorter than $\sim 1$ Gyr (Thomas et  al. 2005).  Therefore,
like early-type galaxies the MW bulge is likely to have formed through
a short series of starbursts triggered by  the coalescence of gas-rich
mergers, when the universe was only a few Gyr old.

\begin{acknowledgements}
We  thank Ortwin Gerhard  for drawing our attention  to the results of
the Immeli et al. investigation.  This  work has been partly funded by
the FONDAP  Center for Astrophysics 15010003 (MZ  and DM), DIPUC (MZ),
and by a Fellowship of the John Simon  Guggenheim Foundation (DM).  DM
acknowledges the  European Commission's ALFA-II programme, through its
funding  of the Latin-American  European  Network for Astrophysics and
Cosmology  (LENAC).  BB  and DM  acknowledge grants  from the CNPq and
Fapesp.  SO  acknowledges the   Italian Ministero  dell'Universit\`a e
della Ricerca   Scientifica e Tecnologica   (MURST) under  the program
'Fasi iniziali   di  evoluzione  dell'alone e del     bulge Galattico'
(Italy).
\end{acknowledgements}



\begin{thebibliography}{}
\bibitem[]{} Allende Prieto, C., Lambert, D.L. \& Asplund, M., 2001, ApJ, 556, L63
\bibitem[]{} Alvarez R. \& Plez B. 1998, A\&A, 330, 1109
\bibitem[]{} Asplund, M., Grevesse, N., Sauval, A.J., Allende Prieto, C. \&
        Kiselman, D. 2004, A\&A, 417, 751
\bibitem[]{} Barbuy, B. et al. 2003, A\&A, 404, 661
\bibitem[]{} Barklem, P.S., Anstee, S.D. \& O'Mara, B.J., 1998, PASA, 15, 336
\bibitem[]{} Barklem, P.S., Piskunov, N.E. \& O'Mara, B. J., 2000, A\&AS, 142, 467
\bibitem[]{} Bensby, T., Feltzing, S. \& Lundstr\"om, I., 2004, A\&A, 421, 155
\bibitem[]{} Carpenter, J.M. 2001, AJ, 121, 2851
\bibitem[]{} Dean, J.F., Warpen, P.R. \& Cousins, A.J. 1978, MNRAS, 183, 569
\bibitem[]{} Fulbright, J.P., McWilliam A. \& Rich R.M., 2006, ApJ, 636, 821
\bibitem[]{} Gratton, R.G., \& Sneden C., 1990, A\&A, 234, 366
\bibitem[]{} Greggio, L. 2005, A\&A, 441, 1055
\bibitem[]{} Grevesse, N. \& Sauval, J. 1998, Space Sci. Rev., 85, 161
\bibitem[]{} Gustafsson, B., Edvardsson, B., Eriksson, K., Mizuno-Wiedner, M., 
J\/orgensen, U.G., Plez, B. in Stellar Atmosphere Modeling
(I.Hubeny, D.Mihalas, K.Werner eds.) ASP Conf. Ser. 2002, Vol. 288, , p. 331
\bibitem[]{} Immeli A., Samland M., Gerhard O. \& Westera P. 2004, A\&A 413, 547
\bibitem[]{} Johansson, S., Litz\'en, U., Lundberg, H. \& Zhang, Z. 2003, ApJ, 584, L107
\bibitem[]{} Kormendy, J., \& Kennicutt, R.C.Jr. 2004, ARA\&A, 42, 603
\bibitem[]{} Kuijken, K. \& Rich, R.M. 2002, AJ, 123, 2054
\bibitem[]{} Martin, W.C. et al. NIST Atomic Database (version 2.0).
        National Institute of Standards and Technology, Gaithersburg, MD. 2002
             (http://physics.nist.gov/asd)
\bibitem[]{} Matteucci, F. \& Greggio L. 1986, A\&A, 154, 279
\bibitem[]{} Matteucci, F., Romano, D., Molaro, P., 1999, A\&A, 341, 458
\bibitem[]{} McWilliam, A. \& Rich, R.M. 1994, ApJ, 91, 749
\bibitem[]{} McWilliam, A. \& Rich, R.M. 2003, in
        Origin and Evolution of the Elements 
        (http://www.ociw.edu/ociw/symposia/series/symposium4/\-proceedings.html)
\bibitem[]{} Minniti D., 1995, AJ, 109, 1663
\bibitem[]{} Ortolani S., et al. 1995, Nature, 377, 701
\bibitem[]{} Ram\'{\i}rez, I. \& Mel\'endez, J., 2005, ApJ, 626, 465
\bibitem[]{} Rich, R.M. \& Origlia, L., 2005, apJ, 634, 1293
\bibitem[]{} Rieke, G.H. \& Lebofsky, M.J., 1985, ApJ 228, 618
\bibitem[]{} Robin, A.C., Reyl\'e, C., Derri\`ere S., \& Picaud S., 2003, A\&A, 409, 523 
\bibitem[]{} Spite, M., 1967, Annales d'Astroph., vol. 30, p.211
\bibitem[]{} Thomas, D., Maraston, C., Bender R., Mendes de Oliveira, C., 2005, ApJ, 621, 673
\bibitem[]{} Udalski, A. et al., 2002, Acta Astron., vol. 52, 217
\bibitem[]{} Zoccali, M. et al. 2004, A\&A, 423, 507
\bibitem[]{} Zoccali, M. et al. 2003, A\&A, 399, 931
\end{thebibliography}
\end{document}